\documentclass[11pt]{article}

\def\be{\begin{equation}}
\def\ee{\end{equation}}
\def\ba{\begin{eqnarray}}
\def\ea{\end{eqnarray}}

\def\bb{\bibitem}

\begin{document}
\title{Hawking radiation for non asymptotically flat dilatonic black holes using gravitational anomaly}
\author{
J.C. Fabris$^{a}$\footnote{e-mail: fabris@pq.cnpq.br}
and G.T. Marques$^{b}$\footnote{e-mail:gtadaiesky@hotmail.com}\\ \\
{\small $^{a}$ Departamento de F\'{\i}sica, Universidade Federal do
Esp\'{\i}rito Santo,}\\ {\small Vit\'oria, 29060-900,
Esp\'{\i}rito Santo, Brazil}\\
{\small $^{b}$ ICIBE-LASIC, Universidade Federal Rural da Amaz\^onia-Brazil,}\\ {\small Bel\'em, 66077-901,
Par\'{a}, Brazil}}
\date{}
\maketitle

\begin{abstract}
The $d$-dimensional scalar field action may be reduced, in the background geometry of a black hole, to a 2-dimensional effective action. In the near horizon region, it appears a gravitational anomaly:
the energy-momentum tensor of the scalar field is not conserved anymore. This anomaly is removed by introducing a term related to the Hawking temperature of the black hole. Even if the
temperature term introduced is not covariant, a gauge transformation may restore the covariance. We apply this method to compute the temperature of the dilatonic non asymptotically
flat black holes. We compare the results with those obtained through other methods.

\end{abstract}
\newpage

\section{Introduction}

After Hawking has discovered that the black holes are not completely black, since they can emit particles with a black body spectrum with a temperature proportional to the surface gravity at
the horizon \cite{haw}, the black hole temperature phenomena has been re-discovered through many different approaches. The black hole particle emission comes from the fact that the vacuum state near the
horizon is not the same as the vacuum state at infinity. Hence, an observer placed in the spatial infinity receives a particle flux, in the high frequency regime, with the same characteristics
of a black body object \cite{birrell, wald, fulling1}.

Many different techniques have been proposed to compute the temperature of a black hole \cite{page,unruh1,unruh2,fulling,haw1,mitra,GL,NSS,BSS,CHM}. A number of these techniques may be used both for
asymptotically flat and asymptotically non-flat black holes. One of the more recent techniques has been proposed in the reference \cite{ang} and employ the concept of gravitational anomaly, which is
characterized by a violation of the energy-momentum conservation law. It is also called Einstein's anomaly.

Essentially, the method based on the gravitational anomaly consists in reducing the scalar field action, originally written in $D$ dimensions, into a two-dimensional effective chiral action
in the vicinity of the horizon ($r=r_H+\delta$). The gravitational anomaly results from chiral theories \cite{zumino,witten,bertlmann1,bertlmann,suzuki} since the symmetry between ingoing and outgoing
modes is broken. The gravitational anomaly method has been applied to static and dynamics background \cite{vagenas1,vagenas2,vagenas3}.

For a charged black hole, besides the gravitational anomaly, there is the gauge anomaly near the horizon, which is connected to the non-conservation of the four-current \cite{zumino,angg}.
In the same way, for a rotating black hole the quantum number $m$, connected with the angular momentum, plays the r\^ole of a charge, and may also lead to the appearance of the
gravitational anomaly \cite{anggr}.

We will apply the gravitational anomaly methodology to a class of non-asymptotically flat (NAF) black holes that appears in the Einstein-Maxwell-Dilaton (EMD) theory. Specifically, we will consider
the linear dilaton case, for which the coupling constant between the scalar field, the Maxwell field and gravity is equal to unity. Since the EMD theory contains a scalar field coupled non-minimally to
the maxwellian field, such theory can be mapped, through a conformal transformation, to the effective string theory. Such connection between EMD theory and string theory is one of the reason for the
interest in such structures.

We will study more in detail three cases concerning the linear dilatonic solutions: the massive and non-massive static solutions \cite{newdil2} and the rotating solution\cite{cgg}.
We will also consider the topological massive $1+2$ dimensional black hole, coming from a Chern-Simon term coupled to gravity \cite{cedric}. Our approach will consist in considering a
non-covariant formulation of the gravitational anomaly, but we will show that the final results are in agreement with a covariant approach for the energy-momentum tensor and charge flux \cite{BS,banerjee1,banerjee2}.

This paper is organized as follows. In next section we revise the non-covariant gravitational anomaly methodology, showing how the Hawking temperature cancels the anomaly in the vicinity of
the horizon. In Section 3, we apply this procedure to linear dilatonic massive and non massive static black holes. For the last case, there is a gauge anomaly besides the gravitational anomaly. The non massive case represents a massless and chargeless black hole, with a non null surface gravity. However, we show that its Hawking temperature is zero in agreement with the result of reference
\cite{ggj}. In section 4, the dilatonic rotating linear case is analysed. In section 5, the topologically massive black hole is studied. In section 6, we present our conclusions. We include
an appendix to show the consistence of the results obtained through the non covariant formulation of the gravitational anomaly with those obtained from the covariant formulation.

\section{Review of the methodology}

We will be working in this paper with the evaluation of the Hawking temperature via gravitational anomaly developped in the references \cite{ang,angg,anggr}. In order to apply this formalism, the action of
a scalar field in a $D$-dimensional space-time background must reduced to an infinite collection of scalar fields in $1+1$ dimensions. The coordinates are $(t,r)$, $r$ indicating the radial direction and $t$ the time coordinate.

The static spherical symmetric $D$-dimensional metric is given by,
\begin{equation}
\label{met}
ds^2 = f(r)dt^2-f(r)^{-1}dr^2-r^2d\Omega^2_{(D-2)}\quad,
\end{equation}
with $f(r)$ admitting a horizon at $f(r=r_H)=0$.

The action of the scalar field in the limit $r\rightarrow r_H$, for the metric (\ref{met}), is given by,
\begin{eqnarray}
S[\varphi]&=&\frac{1}{2}\int d^Dx\sqrt{-g}\varphi\nabla^2\varphi\nonumber\\
          &=&\frac{1}{2}\int d^Dxr^{D-2}\sqrt{\gamma}\varphi\left[\frac{1}{f}\,\partial_t^2\varphi-\frac{1}{r^{D-2}}\partial_r\left(r^{D-2}f\partial_r\varphi\right)-\frac{1}{r^2}\nabla_\Omega\varphi\right]\nonumber\\
\label{Acaoh}&=&\sum_n\frac{r_H^{D-2}}{2}\int dtdr\,\varphi_n\left[\frac{1}{f}\,\partial_t^2\varphi_n-\partial_r\left(f\partial_r\varphi_n\right)\right]\,\,,
\end{eqnarray}
$\gamma$ being the determinant of the angular section $d\Omega^2$ and $\nabla_\Omega$ a collection of angular derivatives. The scalar field $\varphi$ has been expanded in terms of spherical harmonic functions in $D-2$ dimensions labelled by the index $n$, and a constant term that appears in the near-horizon approximation has been discarded since it does not contribute to the equations of motion \cite{kim}. The orthogonality of the spherical harmonic functions has also been used. Hence, near the horizon, the field can be described by an infinite collection of fields living in the space-time ($1+1$) given by the metric,
\begin{equation}
\label{met1}
ds^2 = f(r)dt^2-f(r)^{-1}dr^2\quad.
\end{equation}
We end up with a two-dimensional action, corresponding to an effective theory near the horizon. In this action we can impose the condition the ingoing modes are absent, since
we are interested in the radiation arriving at infinity. This effective near horizon theory will then be connected with a non chiral theory outside
the horizon, which contains both the ingoing and outgoing modes.

The gravitational anomaly to be treated here is represented by the non conservation of the energy-momentum tensor in two dimensions. The energy-momentum tensor refers to the scalar field
living in the space-time determined by the black hole. Hence, even if the space-time may be asymptotically flat, the energy-momentum tensor may be non-null at infinity representing, at this region
the particle flux from the hole. In the region, $r_H < r\leq r_H+\delta$ the anomaly is given by, \cite{ang,witten,bertlmann}:
\begin{equation}
\label{eang}\nabla_\mu T^\mu_{\nu}(r_H)\equiv\Xi_\nu(r)\equiv\frac{1}{\sqrt{-g}}\,\partial_\mu N^\mu_\nu\quad,
\end{equation}
where,
\begin{equation}
\label{ano}
N^\mu_\nu=\frac{1}{96\pi}\,\epsilon^{\beta\mu}\partial_\alpha\Gamma^\alpha_{\nu\beta}\rightarrow N^r_t=\frac{1}{192\pi}\bigg(ff''+f'\,^2\bigg)\quad,
\end{equation}
with $\epsilon^{01}=\epsilon^{tr}=1$ and ($'$) is a derivative with respect to $r$.

For the metric (\ref{met1}), it is easy to see that it is a timelike anomaly, $\Xi_t\neq0$ and $\Xi_r=0$.

In the external region, where $r > r_H + \delta$, the usual conservation law is satisfied, $\nabla_\mu T^\mu_{\nu}(r) = 0$.

The equation (\ref{eang}) must be solved in two regions, $r_H<r\leq r_H+\delta$, where the anomalies appear, and $r>r_H+\delta$, where there is no anomaly.

\begin{itemize}
\item For $r>r_H+\delta$,
\begin{equation}
\label{tmeo}\partial_rT^r_{t}(\infty)=0\rightarrow T^r_{t}(\infty)=a_0\quad.
\end{equation}
\item For $r_H<r\leq r_H+\delta$,
\begin{equation}
\partial_rT^r_{t}(r_H)=\Xi_t\rightarrow T^r_{t}(r_H)=a_H+N^r_t(r)-N^r_t(r_H)\quad.
\end{equation}
\end{itemize}
In the expressions above $a_0$ and $a_H$ are integration constants.

We combine the energy-momentum tensor in the two regions, leading to,
\begin{equation}
T^\mu_\nu=T^\mu_{\nu}(\infty)\Theta_H(r)+T^\mu_{\nu}(r_H)H(r)\quad,
\end{equation}
where $\Theta_H(r)=\Theta(r-r_H-\delta)$ is the step function and $H(r)=1-\Theta_H(r)$ a hat function which is $1$ in the region $r_H<r\leq r_H+\delta$ and zero outside this region.

Through the transformation, $x'=x-\xi\rightarrow\delta g_{\mu\nu}=\nabla_\mu\xi_\nu+\nabla_\nu\xi_\mu$, the anomaly of the effective action (suppressing the ingoing modes as described above) is given by
\begin{eqnarray}
-\delta W&=&\int d^2x\sqrt{-g^{(2)}}\xi^t\nabla_\mu T^\mu_t \nonumber\\
&=&\int d^2x\,\xi^t\bigg\{\partial_r\bigg[N^r_t(r)H(r)\bigg] \nonumber\\
&+&\bigg(a_0-a_H+N^r_t(r_H)\bigg)\delta(r-r_H-\delta)\bigg\}\quad.
\end{eqnarray}
The first term is zero since it can be reduced to a surface term ($\xi^t$ is constant). But, in order to cancel the second term, it is necessary to consider the quantum effects associated to the ingoing modes, given by,
\begin{equation}
\label{ca01} a_0=a_H-N^r_t(r_H)\quad.
\end{equation}

One consequence of the equation (\ref{eang}) is the absence of covariance. In order to restore the covariance, is necessary to add a new energy-momentum tensor \cite{bertlmann1} ( see details in the
appendix) such that,
\begin{equation}
\label{ntme}\tilde{T}^\mu_\nu=T^\mu_\nu+\tilde{N}^\mu_\nu\quad,
\end{equation}
where
\begin{equation}
\tilde{N}^r_t=\frac{1}{192\pi}\bigg(ff''-2(f')^2\bigg)\quad.
\end{equation}

The value of the constant $a_H$ is fixed by imposing that the covariant energy-momentum tensor is zero over the horizon ($\tilde{T}^r_t(r_H)=0$). Following this procedure, we find $a_H$:
\begin{eqnarray}
\tilde{T}^r_t(r_H)&=&T^r_t(r_H)+\tilde{N}^r_t(r_H)=0 \nonumber\\
\label{cah}a_H&=&\frac{1}{96\pi}\,\left(f'|_{r=r_H}\right)^2\quad.
\end{eqnarray}
Remark that $\tilde{N}^r_t(r_H)=-2N^r_t(r_H)$, since $f(r_H)=0$. The equation (\ref{ca01}) may be written as,
\begin{eqnarray*}
a_0&=&a_H-N^r_t(r_H)\\
&=&\frac{1}{48\pi}\,\kappa^2\quad,
\end{eqnarray*}
where $\kappa=\frac{1}{2}\partial_r f|_{r=r_H}$ is the surface gravity. Using now the expression for the temperature of the black hole $T_H=\kappa/2\pi$, we can express the flux of energy-momentum as,
\begin{equation}
a_0=\frac{\pi}{12}\,T_H^2\quad.
\end{equation}

The main point is the that the temperature term is essential to cancel the gravitational anomaly in the near horizon region.
\par
In this section, the method has been exemplified for the case of a static black hole. However, it can be applied as outlined above, essentially without changes, to the case of rotating
black holes, for which the space-time is not static but stationary. The reason is that the method is based on the reduction of the action to an effective two-dimensional action. In two dimensions, the original non-diagonal metric, which is symmetric, can be diagonalized, and the case of the rotating black hole can be recast in the structure developed above \cite{anggr,soda}. This will be
seen in details in sections IV and V.

\section{The NAF Dilatonic black holes}

The action describing the dilatonic field coupled to gravity and to a gauge field $U(1)$ in $3+1$ dimensions is given by,
\begin{equation}
\label{emda}S=\frac{1}{16\pi}\int d^4x\sqrt{-g}\left(R-2\partial_\mu\phi\partial^\mu\phi-e^{-2\alpha\phi}F_{\mu\nu}F^{\mu\nu}\right)\nonumber\,\,,
\end{equation}
where $\phi$ is the dilatonic field, $F_{\mu\nu}$ the $U(1)$ gauge field, and $\alpha$ is a coupling constant. Static, spherically symmetric solutions in four dimensions, which are not necessary asymptotically flat, have been determined
in reference \cite{newdil2}. Among these solutions there is the linear dilatonic case ($\alpha=1$), a static, spherical solution non asymptotically flat solution, described by the following expressions \cite{newdil2}:
\begin{eqnarray}
\label{Mnasp}ds^{2} &=& \frac{(r - b)}{r_{0}}dt^{2} - \frac{r_{0}}{(r - b)}dr^{2} -rr_{0}d\Omega^{2}\quad,\\
\label{cenasp}F^{tr}&=& \frac{Q}{r_{0}^{2}}\,\,,\quad e^{2\phi(r)}=\frac{r}{r_{0}}\quad,
\end{eqnarray}
In these expressions, $b$ are $r_0$ are related to the mass and charge of the black hole:
\begin{equation}
\label{Q}M=\frac{1}{4}\,b\,\,,\quad Q=\sqrt{\frac{1}{2}}\,r_0\quad.
\end{equation}

Near the horizon the effective two-dimensional metric is given by,
\begin{equation}
\label{Mnasp2d}ds^{2}=\frac{(r - b)}{r_{0}}dt^{2} - \frac{r_{0}}{(r - b)}dr^{2}\quad.
\end{equation}
This solution contains a gauge term. As for the pure gravitational system, now there is an anomaly in presence of gauge fields which is given by \cite{angg},
\begin{equation}
\label{ang}\nabla_\mu J^\mu=-\,\frac{e^2}{4\pi\sqrt{-g}}\,\epsilon^{\mu\nu}\partial_\mu A_\nu\quad ,
\end{equation}
where $J^\mu$ is the non-conserved four-current and $A_\nu$ the electromagnetic potential. As it has been described in the pure gravitational case, we are interested in the outgoing modes only.

We consider a charged scalar field living in the spacetime determined by the metric (\ref{Mnasp2d}). An energy-momentum tensor and a current are associated to this
charged scalar field. The equation (\ref{ang}) must be solved in the two regions we are interested in:
\begin{itemize}
\item $r>r_H+\delta$, where the conservation laws are satisfied,
\begin{eqnarray}
\nabla_\mu J^\mu(\infty)&=&\partial_\mu J^\mu(\infty)=\partial_r J^r(\infty)=0\nonumber\, ,\\
\label{cc}J^r(\infty)&=&c_0\quad.
\end{eqnarray}
\item  $r_H<r\leq r_H+\delta$, where the anomaly appears,
\begin{eqnarray}
\partial_r J^r(r_H)&=&-\,\frac{e^2}{4\pi}\,\epsilon^{rt}\partial_r A_t\nonumber\, ,\\
\label{cph}J^r(r_H)&=&c_H+\frac{e^2}{4\pi}\,(A_t(r)-A_t(r_H))\quad.
\end{eqnarray}
\end{itemize}

Under an infinitesimal gauge transformation, after omitting the ingoing modes near the horizon, the effective action is,
\begin{displaymath}
-\delta W= \int d^2x \sqrt{-g_{(2)}}\lambda\nabla_\mu J^\mu_{(2)}\quad,
\end{displaymath}
where $\lambda$ is the gauge parameter. As in the gravitational anomaly case, the current is the sum of the currents in the two regions,
\begin{equation}
J^\mu=J^\mu(\infty)\Theta_H(r)+J^\mu(r_H)H(r)\quad,
\end{equation}
where $\Theta_H(r)=\Theta_H(r-r_H-\delta)$ and $H(r)=1-\Theta_H(r)$.

Using the equations (\ref{cc}) and (\ref{cph}), we find,
\begin{eqnarray*}
\int d^2x \sqrt{-g_{(2)}}\lambda\nabla_\mu J^\mu_{(2)}
&=&\int d^2x \lambda\bigg\{\partial_r\left(\frac{e^2}{4\pi}\,A_t(r)H(r)\right)\\
&+&\left(c_0-c_H+\frac{e^2}{4\pi}\,A_t(r_H)\right)\delta(r-r_H-\delta)\biggr\}\, .
\end{eqnarray*}
The first term is zero classically. But, in order to cancel the second term, it is necessary to take into account the quantum terms:
\begin{equation}
\label{ca}c_0=c_H\,-\,\frac{e^2}{4\pi}\,A_t(r_H)\quad.
\end{equation}

Again, as in the case of the energy-momentum tensor anomaly, the equation (\ref{ang}) does not transform covariantly. In order to restore the covariance, we must add a new current \cite{zumino}:
\begin{equation}
\label{cn}\tilde{J}^\mu=J^\mu+\frac{e^2}{4\pi\sqrt{-g}}\,\epsilon^{\lambda\mu}A_\lambda\quad.
\end{equation}

In order to determine the value of the constant $c_H$, it is required that the current (\ref{cn}) is zero near the horizon. Hence,
\begin{equation}
\label{nc}c_H=-\,\frac{e^2}{4\pi}\,A_t(r_H)\quad.
\end{equation}
Substituting $c_H$ in the equation (\ref{ca}), we find the charge flux,
\begin{equation}
\label{c0}c_0=-\,\frac{e^2}{2\pi}\,A_t(r_H=b)\quad.
\end{equation}

The potential given by the equations (\ref{cenasp}) and (\ref{Q}), is given by,
\begin{equation}
A_t(r_H)=\frac{b}{\sqrt{2}\,r_0}\quad.
\end{equation}
The charge flux is,
\begin{equation}
\label{fcnasp}c_0=-\,\frac{e^2}{2\pi}\frac{b}{\sqrt{2}\,r_0}\quad.
\end{equation}

-{\itshape Gravitational Anomaly}: In presence of a current, even at classical level the energy-momentum tensor is not conserved:
\begin{equation}
\label{tmencc}\nabla_\mu T^\mu\,_\nu=F_{\mu\nu}J^\mu\quad.
\end{equation}
Hence, taking into account all fields present in the action, computing the variation with respect $S[g_{\mu\nu}, A_\mu,\varphi,\sigma]$, taking into account the Ward's identities, we obtain
the following expression for the variation of the energy-momentum tensor after adding the gravitational anomaly term \cite{angg}:
\begin{equation}
\label{eqt1}\nabla_\mu T^\mu_\nu=F_{\mu\nu}J^\mu+A_\nu\nabla_\mu J^\mu+\Xi_\nu\quad,
\end{equation}
The flux of the energy-momentum tensor is independent of the dilation field $\sigma$. It must be stressed that the space-time is static and that only the outgoing modes are taken into account.

The equation (\ref{eqt1}) is solved, for $\nu = t$, in the two relevant regions.
\begin{itemize}
\item In $r>r_H+\delta$,
\begin{eqnarray}
\partial_rT^r_{t}(\infty)&=&F_{r t}J^r(\infty)\nonumber\\
\label{tmeo}T^r_{t}(\infty)&=&a_0+c_0A_{t}(r)\quad.
\end{eqnarray}
\item In $r_H<r\leq r_H+\delta$,
\begin{eqnarray}
\partial_rT^r_{t}(r_H)&=&F_{r t}J^r(r_H)+A_t(r)\partial_r J^r(r_H)+\Xi_t\nonumber\\
T^r_{t}(r_H)&=&a_H \nonumber\\
&+& \int_{r_H}^{r}d\biggr(\frac{e^2}{4\pi}A_t^2(r) + c_0A_t(r)+ N_t^r(r)\biggl)\quad,
\end{eqnarray}
where the equations (\ref{cph}), (\ref{nc}) and (\ref{c0}) have been used.
\end{itemize}
Above, $a_0$ and $a_H$ are integration constants.
\par
Collecting all terms, the total energy-momentum tensor can be written as
\begin{eqnarray}
T^r_t(r) &=& \biggr(\tilde T^r_t(r_H) + \bar T^r_t(r_H)\biggl)H(r - r_h - \delta) \nonumber\\
&+& \biggr(\tilde T^r_t(\infty) + \bar T^r_t(\infty)\biggl)\Theta(r - r_h - \delta),
\end{eqnarray}
where $\tilde T^r_t$ collects the constants contributions and $\bar T^r_t$ contains the terms that are $r$ dependent.

Following the same steps as before, the anomaly part of the effective action is given by,
\begin{eqnarray*}
\int d^2x\sqrt{-g_{(2)}}\xi^t\nabla_\mu T^\mu\,_t
&=&\int d^2x\,\xi^t\biggr\{\nabla_r\biggr[\bar T^r_t(r_H)H(r - r_H - \delta) \nonumber\\
&+& \bar T_t^r(\infty)\left(1+\Theta(r - r_H - \delta)\right)\biggl]\nonumber\\
&+&\left(a_0-a_H - \frac{e^2}{4\pi}\,A_t^2(r_H) + N^r_t(r_H)\right)\delta(r-r_H-\delta)\bigg\}\,\,.
\end{eqnarray*}

To cancel the anomaly, it is required that,
\begin{equation}
\label{ca0} a_0=a_H + \frac{e^2}{4\pi}\,A_t^2(r_H) - N^r_t(r_H)\quad.
\end{equation}

Equation (\ref{eang}) does not transform covariantly. In order to restore the covariance, a new energy-momentum tensor is added \cite{bertlmann1}:
\begin{equation}
\label{ntme}\tilde{T}^\mu_\nu=T^\mu_\nu+\tilde{N}^\mu_\nu\quad,
\end{equation}
where
\begin{equation}
\tilde{N}^r_t=\frac{1}{192\pi}\bigg(ff''-2(f')^2\bigg)\quad.
\end{equation}
Imposing that the covariant energy-momentum tensor (\ref{ntme}) be zero on the horizon, the flux of the energy-momentum tensor (\ref{ca0}) becomes,
\begin{equation}
\label{tmenasp}a_H = - \tilde N^r_t(r_H).
\end{equation}
Hence,
\begin{eqnarray}
 a_0 &=& \frac{e^2}{4\pi}\,A_t^2(r_H) - \tilde N^r_t(r_H) - N^r_t(r_H)\nonumber\\
\label{ca0bis} &=&\frac{e^2b^2}{8\pi r_0^2} + \frac{1}{192\pi}\frac{1}{r_0^2} \quad,
\end{eqnarray}
implying the temperature,
\begin{eqnarray}
T_H = \frac{1}{4\pi r_0},
\end{eqnarray}
confirming the temperature found in \cite{ggj}.

\subsection{Case $b=0$}

This case represents the dilatonic vacuum $|0,r_0>$ which is given by the metric,
\begin{equation}
\label{mvd}ds^2=\frac{r}{r_0}\,dt^2-\frac{r_0}{r}\,dr^2-rr_0d\Omega^2\quad,
\end{equation}
which also describes the geometry in the asymptotic limit ($r\rightarrow\infty$) of the linear dilatonic case with
$b\neq0$. Remark that, in this case, the horizon coincides with the singularity. However, since the singularity is lightlike, it is still a true
black hole: the signal emitted from the singularity reach an external observer in an infinite time. Hence, it is a black hole structure, not a naked singularity.

Computing the temperature via surface gravity for the metric (\ref{mvd}), we find,
\begin{displaymath}
T_H=\frac{1}{4\pi r_0}\quad.
\end{displaymath}

Using the coordinate transformation $x=\ln(r/r_0)$ and $\tau=t/r_0$ \cite{ggj}, the metric takes the form,
\begin{equation}
\label{cm}ds^2=\Sigma^2\left(d\tau^2-dx^2-d\Omega^2\right)\quad,
\end{equation}
where $\Sigma=r_0e^{x/2}=rr_0$ is a conformal factor.

The linear dilaton vacuum is represented by a metric conformal to the product $M_2XS_2$, a two-dimensional Minkowski space-time, and a unitary two sphere.

The action of the scalar field takes the form,
\begin{displaymath}
S=\int d^4x\sqrt{-g}\psi\nabla^2\psi\,\,.
\end{displaymath}
Redefining the field $\psi=\Sigma^{-1}\Psi$,the action for the metric (\ref{cm}) takes the form,
\begin{equation}
S =\int d^4x\sin\theta\Psi\left(\partial^2_{\tau} - \partial^2_x - \nabla^2_{\Omega} + \frac{1}{4}\right)\Psi.
\end{equation}
This action represents free field, as shown in reference \cite{ggj}, where there is no anomaly to be treated. Hence, there is no particle flux in this case.
In fact, using expression (\ref{ano}), we find that $N^r_t$ is, for this case, constant, and no anomaly appears. This result agrees with that exhibit in reference \cite{ggj}, where the
method of transmission/refection coefficients has been used.
\par
The vacuum linear dilatonic case corresponds to a massless, but charged
 black hole. Moreover, as stated above, the horizon coincides with the singularity, but it is not a naked singularity since any signal sent from the horizon takes an infinite time to reach any external observer, as it happens with the usual event horizon. Due to all these properties, it constitutes a very exotic object, and it is not surprising that its thermodynamics is somehow
 ill defined, with a discrepancy for the temperature obtained from the surface gravity and from other methods, like the transmission/reflection coefficients and the anomaly method.
In some sense, the vacuum linear dilatonic case has some similarity with extremal black holes from the thermodynamic point of view, which may explain the discrepancy between the surface gravity and the temperature obtained by the anomaly method. A computation using the Bogolioubov coefficients for the Reissner-Nordstr\"om extremal black hole has shown that perhaps
the thermodynamics of these objects are ill defined, see reference \cite{extremo}, and references therein. In the reference \cite{oleg}, it has also been shown that the entropy of
the extremal black hole may be zero, even  when the horizon surface is infinite, revealing a discrepancy from the usual black hole entropy law. The existence of all these discrepancies
reveals that the results obtained here for the vacuum linear dilatonic case is not so exceptional.

\section{Linear dilaton black holes with rotation}

In this section, we will consider again the linear dilatonic theory \cite{cgg}, $\alpha=1$. In this case, the EMD theory is reduced to the heterotic string theory in 4D.
In the linear dilatonic case, the EMD theory is a sector of a more general theory: the Einstein-Maxwell-dilatonic-axionic (EMDA) theory. This theory, with $\alpha=1$,
admits a formulation in terms of the $\sigma$ model. In this way, it can be found a metric with rotation, which is a generalization of
the static metric with $\alpha=1$ studied in the previous section.

The metric \cite{cgg},
\begin{equation}
\label{mdlr}
ds^2=\frac{\Delta(e)}{r_0r}\,dt^2+2a\sin^2\theta dtd\varphi-\frac{r_0r}{\Delta}\,dr^2-r_0rd\Omega^2\quad,
\end{equation}
represents the geometry in the linear dilaton case with rotation of the theory EMDA. In (\ref{mdlr}), we have
\begin{eqnarray*}
\Delta(e)&=&(r-r^{(e)}_+)(r-r^{(e)}_-)=r^2-2Mr+a^2\cos^2\theta\quad,\nonumber\\
r^{(e)}_\pm &=& M\pm\sqrt{M^2-a^2\cos^2\theta}\quad,\nonumber\\
\Delta&=&(r-r_+)(r-r_-)=r^2-2Mr+a^2\quad,\nonumber\\
r_\pm &=& M\pm\sqrt{M^2-a^2}\quad,
\end{eqnarray*}
and from (\ref{mdlr}),
\begin{eqnarray*}
\Omega_h&=&\frac{g_{t\varphi}}{g_{\varphi\varphi}}|_{r=r_+}=\frac{a}{r_0r_+}\quad,\nonumber\\
A&=&\frac{1}{\sqrt{2}}\left[\frac{(r^2+a\cos^2\theta)}{rr_0}\,dt+a\sin^2\theta d\varphi\right]\quad,\nonumber\\
\label{r1}\kappa&=&\frac{r_+-r_-}{2r_0r_+}\quad,
\end{eqnarray*}
where $\Omega_h$ is the angular velocity over the horizon, $r_\pm$ the external and internal horizon, $r^{(e)}_+$ the stationary region limit, $M=b/2$ the mass of the black hole, determined by the quasi-local formalism, $A$ the four potential and $\kappa$ the surface gravity (\ref{mdlr}).
Rotating black holes have a charge represented by the azimuthal quantum number
$m$ \cite{anggr}.

Let us consider a charged scalar field in the space-time defined by the metric (\ref{mdlr}). The action of the charged scalar field, after expanded in the spherical harmonics and in the
near horizon approximation, $r=r_++\delta$, is given by,
\begin{eqnarray}
S&=&\int d^4x\sqrt{|g|}\phi^*\nabla^2\phi.
\end{eqnarray}
In this expression, we have
\begin{eqnarray*}
\nabla^2\phi&=&g^{\mu\nu}{\cal D}_\mu{\cal D}_\nu\phi=g^{\mu\nu}(\nabla_\mu-ieA_\mu)(\nabla_\nu-ieA_\nu)\phi\\
&=&-\frac{1}{rr_0}\,\partial_r\Delta\partial_r\phi-\frac{1}{rr_0}\,\nabla_\Omega\phi+\frac{rr_0}{\Delta}\,\partial_t^2\phi+\frac{2i}{\Delta}\left[am-\frac{\left(r^2+a^2\right)}{\sqrt{2}}\,e\right]\partial_t\phi\\
&-&\frac{1}{rr_0\Delta}\left\{a^2m^2-\frac{2\left(r^2+a^2-\Delta\right)a}{\sqrt{2}}\,me+\frac{e^2}{2}\left[\left(r^2+a^2\right)^2-\Delta a^2\sin^2\theta\right]\right\}\,\phi\quad.
\end{eqnarray*}
To obtain the expression above, the angular part has been decomposed as follows:
\begin{displaymath}
\phi = \phi(r,t)_{n,m}Y(\theta,\varphi)_{n,m}\quad.
\end{displaymath}

The action takes then the form,
\begin{eqnarray*}
S
&=&\int d^4xrr_0\sin\theta\phi^*\left\{-\frac{1}{rr_0}\,\partial_r(\Delta\partial_r\phi)-\frac{1}{rr_0}\,\nabla_\Omega\phi+\frac{rr_0}{\Delta}\,\partial_t^2\phi\right.\\
&+&\frac{2i}{\Delta}\left[am-\frac{\left(r^2+a^2\right)}{\sqrt{2}}\,e\right]\partial_t\phi-\frac{1}{rr_0\Delta}\left\{a^2m^2\right.\\
&-&\left.\left.\frac{2\left(r^2+a^2-\Delta\right)a}{\sqrt{2}}\,me+\frac{e^2}{2}\left[\left(r^2+a^2\right)^2-\Delta a^2\sin^2\theta\right]\right\}\,\phi\right\}\quad.
\end{eqnarray*}

Employing the tortoise radial coordinate $r*$ defined as,
\begin{eqnarray*}
dr^*
 &=&\frac{rr_0}{\Delta}\,dr\quad,\\
\end{eqnarray*}
the action becomes,
\begin{eqnarray*}
S
&=&\int dtdr^*d\theta d\varphi\sin\theta\phi^*\left\{-\,\partial_{r^*}(rr_0\partial_{r^*}\phi)-\frac{\Delta}{rr_0}\,\nabla_\Omega\phi+rr_0\,\partial_t^2\phi\right.\\
&+&2i\left[am-\frac{\left(r^2+a^2\right)}{\sqrt{2}}\,e\right]\partial_t\phi-\frac{1}{rr_0}\left\{a^2m^2\right.\\
&-&\left.\left.\frac{2\left(r^2+a^2-\Delta\right)a}{\sqrt{2}}\,me+\frac{e^2}{2}\left[\left(r^2+a^2\right)^2-\Delta a^2\sin^2\theta\right]\right\}\,\phi\right\}\quad.
\end{eqnarray*}

Near the horizon, $r\rightarrow r_+\Rightarrow\Delta\rightarrow0$, the expression
for the action simplifies to:
\begin{eqnarray*}
S
&\approx&\int dtdrd\theta d\varphi\,rr_0\sin\theta\phi^*\left[-\,\frac{1}{rr_0}\partial_{r}\Delta\partial_{r}\phi+\frac{rr_0}{\Delta}\,\left(\partial_t+\frac{iam}{rr_0}-\frac{i\left(r^2+a^2\right)e}{rr_0\sqrt{2}}\right)^2\phi\right]\quad.
\end{eqnarray*}
After the expansion in spherical harmonics, the action reads,
\begin{eqnarray*}
S&=&\int dtdr\,rr_0\phi(r,t)^*_{n',m'}\left[-\,\frac{1}{rr_0}\partial_{r}(\Delta\partial_{r}\phi)+\frac{rr_0}{\Delta}\,\left(\partial_t+\frac{iam}{rr_0}-\frac{i\left(r^2+a^2\right)e}{rr_0\sqrt{2}}\right)^2\phi(r,t)_{n,m}\right]\\
&\times&\int d\varphi\int d\theta\sin\theta Y_{n,m}(\theta,\varphi)Y_{n',m'}(\theta,\varphi)^*\\
&=&\int dtdr\,rr_0\phi(r,t)^*_{n,m}\left[-\,\frac{1}{rr_0}\partial_{r}(\Delta\partial_{r}\phi)+\frac{rr_0}{\Delta}\,\left(\partial_t+\frac{iam}{rr_0}-\frac{i\left(r^2+a^2\right)e}{rr_0\sqrt{2}}\right)^2\phi(r,t)_{n,m}\right]\,.
\end{eqnarray*}
The final expression for the action shows that, near the horizon, we have an effective two dimensional metric,
\begin{displaymath}
ds^2=f(r)dt^2-f(r)^{-1}dr^2\quad,
\end{displaymath}
and from now on we can proceed as in the static case.

From the considerations made above and from the quantum theory of a charged scalar field, we have
\begin{eqnarray}
g_{tt}&=&-g_{rr}^{-1}=f(r)=\frac{\Delta}{r_0r}\quad,\\
{\cal A}_t(r)&=&\frac{e}{rr_0}\frac{(r^2+a^2)}{\sqrt{2}}-\,\frac{ma}{r_0r}=eA_t^{(1)}+mA_t^{(2)}\quad,
\end{eqnarray}
where $m$ plays the of the topological $U(1)$ charge for the two-dimensional field $\phi_{lm}$. As it happens in the Kerr-Newman metric
\cite{anggr} the potential is the sum of two terms: the first term comes from the electric charge $r_0$, from the black hole electric field,
and the second term is the gauge potential induced by the metric, associated to the symmetry of the background metric (\ref{mdlr}).

The procedure to calculate the flux via anomaly is identical to the charged black hole, presented in the previous section.

Consider the conserved current:
\begin{displaymath}
\partial_rJ^r(\infty)=0\,,\,\,J^r(\infty)=c_0\quad.
\end{displaymath}
For the anomaly part, we evaluate the two potentials, finding
\begin{eqnarray}
A_t^{(1)}(r_H)&=&\frac{(r_H^2+a^2)}{r_Hr_0\sqrt{2}}\quad,\\
A_t^{(2)}(r_H)&=&-\,\frac{a}{r_0r_H}\quad,
\end{eqnarray}
which are the potentials related to the charge $r_0$ and the metric induced angular momentum $a$, respectively.

From equation (\ref{c0}), the charge flux generated by the charge $r_0$, and angular momentum flux are,
\begin{eqnarray}
\label{fcdlr*}c^{(1)}_0&=&-\frac{e}{2\pi}{\cal A}_t(r_H)=\frac{e}{2\pi}\left[m\Omega_H-\frac{e}{r_+r_0}\frac{(r_+^2+a^2)}{\sqrt{2}}\right]\quad,\\
\label{fcdlr}c^{(2)}_0&=&-\frac{m}{2\pi}{\cal A}_t(r_H)=\frac{m}{2\pi}\left[m\Omega_H-\frac{e}{r_+r_0}\frac{(r_+^2+a^2)}{\sqrt{2}}\right]\quad,
\end{eqnarray}
respectively

Finally, the equations for the anomalies given by (\ref{eqt1})for the energy-momentum tensor in the region $r\in[r_+,r_++\delta]$, near the horizon, is given by
\begin{equation}
\label{eqtdlr}\partial_r T^r_t={\cal F}_{rt}{\cal J}^r+{\cal A}_t\partial_r{\cal J}^r+\Xi_t\quad,
\end{equation}
where ${\cal F}_{rt}=\partial_r{\cal A}_t$. ${\cal J}^r$, as defined before, satisfies $\partial_r{\cal J}^r=\frac{1}{4\pi}\partial_r{\cal A}_t$. Applying the same method of the static case,
the flux of the energy-momentum tensor is determined as,
\begin{eqnarray}
a_0&=&\frac{{\cal A}_t^2(r_H)}{4\pi}+\frac{\pi}{12}\,T_H^2\nonumber\\
\label{tmedlr}&=&\frac{1}{4\pi}\left[m\Omega_H-\frac{e}{r_+r_0}\frac{(r_+^2+a^2)}{\sqrt{2}}\right]^2+\frac{\pi}{12}\left(\frac{r_+-r_-}{4\pi r_0r_+}\right)^2\, .
\end{eqnarray}
The temperature is in agreement with the surface gravity given in the reference \cite{cgg}.

If we set $a=0$, the equations for the flux of charge (\ref{fcdlr*}) and of the energy-momentum tensor (\ref{tmedlr}) become equations (\ref{fcnasp}) for the charge flux and (\ref{ca0bis}) for the energy-momentum tensor, for the static case studied in the previous section.

\section{Topologically massive black holes}

In 1982, Deser, Jackiw and Templeton \cite{gtm} have introduced the Topological Massive Gravity, which is essentially the 3D General Relativity coupled to the Chern-Simons term.
The action of this theory is,
\begin{equation}
S=\frac{1}{16\pi G}\,\int d^3x \left[\sqrt{-g}\,R+\frac{1}{\mu}\,\epsilon^{\lambda\nu\sigma}\Gamma^{\beta}_{\lambda\upsilon}\left(\partial_\nu\Gamma^\upsilon_{\beta\sigma}+\frac{2}{3}\,\Gamma^\upsilon_{\nu\tau}\Gamma^\tau_{\sigma\beta}\right)\right]\quad,
\end{equation}
where $R$ is the Ricci scalar, $\Gamma$ is Christoffel symbol, $G$ is the Newton gravitation constant and $\mu$ the Chern-Simons coupling constant. The Chern-Simons term is also called the topological term, since it does not depend explicitly of the metric. On the other hand, this term makes appear a new dynamic degree of freedom: a particle of spin 2 and
mass $\mu$.

A solution of this theory is given by the metric \cite{cedric},
\begin{equation}
\label{mtgm}ds^2=3dt^2-(4\rho+6\omega)dtd\varphi+\frac{d\rho^2}{\rho^2-\rho^2_0}+r^2d\varphi^2\quad,
\end{equation}
where $r^2=\rho^2+4\omega\rho+3\omega^2+\rho_0^2 /3$, $\rho_\pm=\pm\rho_0$ are the internal and external horizons events, $\Omega_H=3/(2\rho_0+3\omega)$ is the angular velocity
over the horizon, $g=-1$ is the metric determinant and $\omega$ is an integration constant of the theory. The surface gravity is given by $\kappa = \sqrt{3}\rho_0/(2\rho_0+3\omega)$.

In the same way as in the previous section, we construct the action of the charged scalar field. Near the horizon is given by,
\begin{eqnarray}
S&=&\int d^3x\sqrt{-g}\phi^*\nabla^2\phi\nonumber\\
&\approx&\int d^2x\,\frac{\left(2\rho_0+3\omega\right)}{\sqrt{3}}\phi^*_m\left\{\partial_\rho\left[\left(\frac{\rho^2-\rho^2_0}{2\rho_0+3\omega}\right)\sqrt{3}\partial_\rho\phi_m\right]\right.\nonumber \\
&&\left.-\frac{1}{\sqrt{3}}\left(\frac{2\rho_0+3\omega}{\rho^2-\rho^2_0}\right)\left[\partial_t\phi_m+im\,\frac{3\phi_m}{(2\rho_0+3\omega)}\right]^2\right\}\quad.
\end{eqnarray}

The expression above for the action of the scalar field defines the effective two-dimensional metric in the quasi-global coordinates. Hence, we have,
\begin{eqnarray*}
g_{tt}&=&-g_{\rho\rho}^{-1}=f(\rho)=\left(\frac{\rho^2-\rho^2_0}{2\rho_0+3\omega}\right)\sqrt{3}\quad,\\
A_t(\rho)&=&-\,\frac{3}{(2\rho+3\omega)}\quad.
\end{eqnarray*}

Following the same steps of the previous case for the rotating black hole, the charge flux is now represented by the azimuthal quantum number $m$, the charge flux and the flux of the energy-momentum tensor are
\begin{eqnarray}
a_0&=&\frac{m^2}{4\pi}\,A_t^2(\rho_H)+\frac{\pi}{12}\,T_H^2\nonumber\\
\label{ftmetgm}&=&\frac{m^2\Omega_H^2}{4\pi}+\frac{\pi}{12}\left(\frac{\sqrt{3}\rho_0}{2\pi(3\omega+2\rho_0)}\right)^2\quad.
\end{eqnarray}
The first term in the right hand side represents the chemical potential of the charge flux while the second one leads to the temperature
\begin{equation}
T_H = \frac{\sqrt{3}\rho_0}{2\pi(3\omega+2\rho_0)},
\end{equation}
in agreement with the expression obtained from the surface gravity. A similar computation for this case has been made in reference \cite{grego}, leading to the same results.
As in all cases studied before, the charge flux represents a chemical potential, in agreement with the black hole thermodynamics.

\section{Conclusion}

The Hawking radiation is given by a Planck distribution,
characteristic of a black body, with chemical potentials for the azimuthal angular momentum $m$, and for the electric charge $e$, of the fields irradiated by the black hole.
For the bosonic spectrum, the distribution reads,
\begin{equation}
N_{e,m}=\frac{1}{e^{\beta(\omega-e\Phi_H-m\Omega_H)}-1}\quad,
\end{equation}
where $\Omega_H$ is the angular velocity over the horizon, 	 and $\beta$ is the inverse of the temperature.
The computation of the Hawking temperature can be made through many different techniques: surface gravity, Bogolioubov coefficients for quantum modes, metric euclideanization,
transmission and reflection coefficients. In general, it is expected that all these methods lead to the same result.

However, such complete equivalence in the computation of the Hawking temperature through different methods may not happen in two cases: extremal black holes and non-asymptotically flat
black holes. One example has been given in reference \cite{ggj}, where the surface gravity gives a finite temperature, but the transmission/reflection coefficients indicate a null result.
In the present work we have exploited the method of gravitational anomaly to compute the temperature of non-asymptotically flat black holes. This anomaly is essentially
related to the non-conservation of the energy-momentum tensor in the quantum effective action.

We have shown the applicability of the method of calculation of the Hawking flux via the cancellation of the anomalies that appear near the event horizon, for non asymptotically flat dilatonic black holes and for topologically massive black hole in ($1+2$) dimensions. The temperature computed through the anomaly method agrees with the surface gravity expression,
showing, in general, the robustness of the results. The only exception is the dilatonic vacuum case, where the anomaly method gives a zero temperature, while the surface
gravity is finite. But, in this last case, there is an agreement with the computation using transmission/reflection coefficients, see reference \cite{ggj}.

We can speculate if such methodology may shed light also in the entropy computation. We hope to address this problem in future works.
\vspace{0.5cm}
\par
\noindent
{\bf Acknowledgements:} J.C.F. thanks CNPq and FAPES (Brazil) for partial financial support. G.T.M. thanks CNPq and FAPESPA (Brazil) for partial financial support. We thank G\'erard Cl\'ement for
valuable remarks on this work.
\newpage

{\bf Appendix - The covariant anomaly}
\vspace{0.5cm}

The anomalies treated in this paper obey strong consistency conditions. For this reason, they are called {\it consistent anomalies}.
This formulation is based on equations (\ref{eang}) and (\ref{ang}) which do not transform covariantly under coordinate transformation. For this reason, the new energy-momentum tensor (\ref{ntme}) and currents (\ref{cn}) are defined, leading to new anomaly equations, which are now invariant under coordinate and gauge transformations. Hence, we will perform an analysis of
covariant equations for the gauge and gravitational anomaly for the metric (\ref{Mnasp2d}).

-{\itshape Gauge covariant anomaly}: Substituting equation (\ref{ang}) in equation (\ref{cn}), we obtain,
\begin{equation}
\label{ancc}\nabla_\mu\tilde{J}^\mu=-\frac{e^2}{4\pi\sqrt{-g}}\epsilon^{\mu\nu}F_{\mu\nu}\quad,
\end{equation}
which is now gauge invariant.

For the metric (\ref{Mnasp2d}) the equations (\ref{ancc}) can be integrated , resulting in
\begin{eqnarray}
\label{j}\partial_r\tilde{J}^r(r)&=&\frac{e^2}{2\pi}\partial_rA_t(r)\\
\tilde{J}^r(\infty)-\tilde{J}^r(r_H)&=&\frac{e^2}{2\pi}(A_t(\infty)-A_t(r_H))\nonumber\quad.
\end{eqnarray}

Imposing that $\tilde{J}^r(r_H)$ is zero over the horizon and that $A_t(\infty)$ is not zero at infinity, since the metric is not asymptotically flat at the infinity, the charge flux becomes,
\begin{equation}
\label{fc}F_c=\tilde{J}^r(\infty)-\frac{e^2}{2\pi}A_t(\infty)=-\frac{e^2}{2\pi}A_t(r_H)=-\frac{e^2}{2\pi}\frac{b}{\sqrt{2}\,r_0}\quad.
\end{equation}
This confirms the universality of the expression for the charge flux, in the asymptotic limit ($r\rightarrow\infty$), via anomalies.

-{\itshape Covariant gravitational anomaly}: The covariant gravitational anomaly is given by the equation \cite{bertlmann1,banerjee1,banerjee2},
\begin{eqnarray}
\nabla_\mu\tilde{T}^\mu_\nu&=&\frac{1}{96\pi}\epsilon_{\mu\nu}\partial^\mu R\quad,\nonumber\\
\partial_r\tilde{T}^r_t&=&\frac{1}{96\pi}f(r)\partial^3_rf(r)=\partial_r\bar{N}^r_t\quad,
\end{eqnarray}
where $R$ is the Ricci scalar and
\begin{displaymath}
\bar{N}^r_t=\frac{1}{96\pi}\left(ff''-\frac{f'^2}{2}\right)\quad,
\end{displaymath}
where $'=d/dr$.

The Ward's identities for the covariant anomaly case is given by \cite{BS,banerjee1}
\begin{equation}
\nabla_\mu\tilde{T}^\mu_\nu=F_{\mu\nu}\tilde{J}^\mu+\partial_\mu\bar{N}^\mu_\nu\quad.
\end{equation}

Using equation (\ref{j}), we have
\begin{eqnarray}
\partial_r\tilde{T}^r_t(r)&=&\partial_rA_t(r)\left(\frac{e^2}{2\pi}\,A_t(r)+d_0\right)+\partial_r\bar{N}^r_t(r)\nonumber\\
\label{fme}\tilde{T}^r_t(\infty)-\tilde{T}^r_t(r_H)&=&\int_{r_H}^{\infty}d\left(\frac{e^2}{4\pi}\,A_t(r)^2+d_0A_t(r)+\bar{N}^r_t(r)\right)\quad.
\end{eqnarray}
Through equation (\ref{j}) it is possible to obtain the value of the constant $d_0$ which is $-e^2\,A_t(r_H)/2\pi$. Hence, after integration of the right hand side of the equation (\ref{fme}), we find
\begin{eqnarray*}
\tilde{T}^r_t(\infty)&=&\frac{e^2}{4\pi}A_t(\infty)^2-\frac{e^2}{2\pi}A_t(\infty)A_t(r_H)+\bar{N}^r_t(\infty)\\
&+&\frac{e^2}{4\pi}A_t(r_H)^2-\bar{N}^r_t(r_H)\quad,
\end{eqnarray*}
where it was used the condition that $\tilde{T}^r_t$ is zero over the horizon. Since the metric (\ref{Mnasp2d}) is not asymptotically flat the term $\bar{N}^r_t(\infty)$ is not zero.
In this way, a flux of energy and momentum at infinity is defined:
\begin{eqnarray}
F_{me}&=&{T}^r_t(\infty)-\frac{e^2}{4\pi}A_t(\infty)^2+\frac{e^2}{2\pi}A_t(\infty)A_t(r_H)-\bar{N}^r_t(\infty)\nonumber\\
&=&\frac{e^2}{4\pi}\,A_t(r_H)^2-\bar{N}^r_t(r_H)=\frac{e^2b^2}{8\pi r_0^2}+\frac{\pi}{12}\left(\frac{1}{4\pi r_0}\right)^2\quad.
\end{eqnarray}
This confirms the equation (\ref{ca0bis}), showing the universality of the method of calculation of the Hawking temperature via
anomalies. Hence, the covariant anomaly method allows also to cancel the anomalies near the horizon, as shown in references \cite{banerjee1,banerjee2,pp}.

\end{document}